# Graphene Quantum Dot as a Probe for DNA Nucleobase Detection: A First Principle Study

Manoj Kumar[1, a)], Neetika Thakur[1] and Munish Sharma[1]

[1]*Department of Physics, School of Basic and Applied Sciences Maharaja Agrasen University, Baddi,, Himachal Pradesh, 174103*

a)Corresponding author: kumarmanu830@gmail.com

**Abstract.** In this study, structural stability, electronic, optical and vibrational properties of DNA nucleobase adsorbed Graphene Quantum Dot (GQD) has been investigated using density functional theory. Based on state-of-art electronic structure calculations, we predict order of GQD sensitivity for DNA nucleobase as Thymine > Cytosine > Guanine > Adenine. An interaction of GQD with DNA nucleobase leads to modulation in electronic energy gap. Our calculated UV/vis and IR vibrational spectra show unique spectral band that can be used fingerprints in next generation DNA sequencing diagnostics.

## INTRODUTCION

Graphene quantum dot (GQD) is the recent addition to the low dimensional material family derived from 2D Graphene [1, 2]. The QDs derived from other 2D materials such as transition metal di-chalcogenides, h-BN, $HfS_2$, $VSe_2$, black phosphorene etc. have unique intrinsic electronic and optical properties, but exhibit low stability, not easily to be functionalized, dispersible in aqueous solution and have very low photoluminescence quantum yield. The pristine GQD offer its candidature to overcome the shortcomings in QD prepared from other 2D materials. GQD has becoming multifunctional material for vide range of applications in Light Emitting Diodes [3, 4], solar cells, batteries, flash memory devices [5], Photovoltaic devices [6] etc. Although, there are numerous studies on the applications of GQD for biomedical applications [7, 8] but there are lack of investigation to explore GQD capabilities towards DNA bio-sensing devices. Metal oxides are being used for sensing applications [[9], but they require high temperature for operation and do not have good sensitivity. GQD offers high surface-to-volume ratio and semiconducting properties which make it suitable candidature to fulfill material-dependent shortcomings for bio-sensing applications.

In this paper, we present the study of interaction of four DNA nucleobases namely Adenine (A), Thymine (T), Cytosine (C) and Guanine (G) with GQD using first principles electronic structure investigations. Effort has been to get insight about electronic properties, UV/vis spectroscopic analysis and IR-vibrational spectroscopy to gauge the scope of GQD for detecting DNA nucleobases.

## SIMULATION DETAILS

We performed a first principle calculations within the framework of DFT using SIESTA code [10]. Electron–ion interactions were treated using well tested Troullier Martin, norm conserving, relativistic pseudo-potential in fully separable Kleinman and Bylander form. PBE functional has been used to account exchange and correlation energies. The double-zeta polarized (DZP) numerical atomic orbital basis set has been used with confinement energy 30 meV. A mesh cutoff of 200 Ry has been used for reciprocal space expansion of the total charge density. A sufficient vacuum size of ~25 Å has been used along all the direction.

The structures were fully optimized in ORCA program. An RIJCOSX method is used to accelerate the SCF calculations. The coulomb exchange term were treated within def2/J auxiliary basis set with valance double-zeta polarization function (def2-SVP) for vibrational spectra calculations. The molecular vibrational spectra have been computed with PBE0 functional. A time-dependent-DFT approach (TDFT) has been employed to get UV/vis spectra with computationally expensive 150 excited states.

## RESULT AND DISCUSSION

### Structural and Electronic Properties

We took a pristine GQD of 36 atoms having Hydrogen terminations. To get optimized geometry, nucleobase has been placed parallel to GQD (Figure 1). The in-plane orientation of nucleobase is reported in past literature too[11, 12].

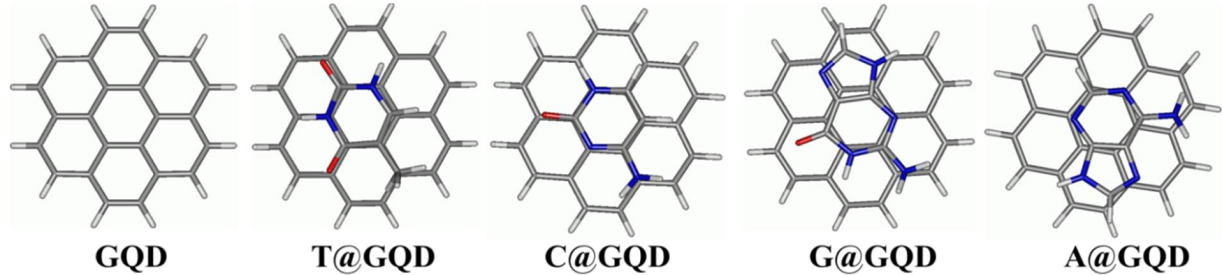

**FIGURE 1.** Stick model (Top view) of optimized structures of graphene quantum dot (GQD) and DNA nucleobases (Thymine, Cytocine, Guanine and Adenine) with GQD

To gauge the structural stability of equilibrium geometries binding energy has been calculated as follows:

$$E_b = E_{nucleobase@GQD} - E_{GQD} - E_{nucleobase} \qquad 1)$$

Where first term represent nucleobase with GQD, second and third term represent total energy of isolated GQD and nucleobase respectively. The obtained binding energy and the optimum height with respect to monolayer have been tabulated in table 1.

**TABLE 1.** Calculated intra-planer distance between nucleobase and GQD, $R_{nucleobase@GQD}$; binding energy, $E_b$ and energy gap for pristine GQD and composite System.

| System | $R_{nucleobase@GQD}$ (Å) | $E_b$ (eV) | Band Gap (eV) |
|---|---|---|---|
| GQD | - | - | 2.75, 3.10 [13] |
| A@GQD | 3.36 | -0.115 | 2.69 |
| T@GQD | 2.98 | -0.173 | 2.74 |
| C@GQD | 3.38 | -0.157 | 2.69 |
| G@GQD | 3.52 | -0.137 | 2.55 |

The negative magnitude of $E_b$ indicates the physisorption of nucleobase on GQD. The magnitude of the calculated binding energy exhibits the following order; Thymine > Cytosine > Guanine > Adenine. We find different optimum height of nucleobase on GQD. Note that Adenine and Cytosine exhibit very close optimum height with difference in corresponding binding energy of 42 meV.

The electronic structure of pristine GQD exhibit wide band gap semiconducting nature with band gap of 2.75 eV which is underestimated by 0.35 eV as compared to reported value of 3.10 eV [13]. This is attributed to inherent discrepancy of DFT to predict the energy gap due to unknown form of exchange correlation functional. The projected density of states (PDOS) of pristine GQD has major contribution to HOMO and LUMO from the C atoms. Nucleobases introduces new energy levels in the vicinity of Fermi Level. In case of nucleobase@GQD LUMO is dominated by the C atoms while the HOMO is modified due to hybridization of N and O p-orbitals with C-2p orbital of GQD (figure not presented here).

## UV/Vis Spectra

UV/Vis spectroscopy is a widely used analytical technique to quantitatively characterize chemical compounds and bio-molecular system. UV/Vis region is region where the photon energy is appropriate to probe the chemical bonds. Figure 2 shows the optical absorption spectra for pristine and nucleobase adsorbed GQD. A characteristic π–π* absorption band of GQD has been captured in absorption spectra with broad peak near 35000 cm$^{-1}$ (i.e between 250 and 300 nm) in all types of considered systems. The absorption bands peak near 35000 cm$^{-1}$ (276.3 nm) is in close agreement with the observed experimental value of 250 nm [13] and 260 nm [2]. Distinct absorption spectra have been observed for different nucleobases as compared to pristine GQD. We found a red shift in absorption spectra as far as different nucleobase is concerned. The π–π* absorption bands confirms the appearance of new energy levels introduced by nucleobases as observed in PDOS analysis. Therefore, it can be expected that the luminescence in the longer wavelength region is caused by extrinsic emissions from the new energy states. Therefore, absorption spectra could serve as a fingerprint the individual nucleobase.

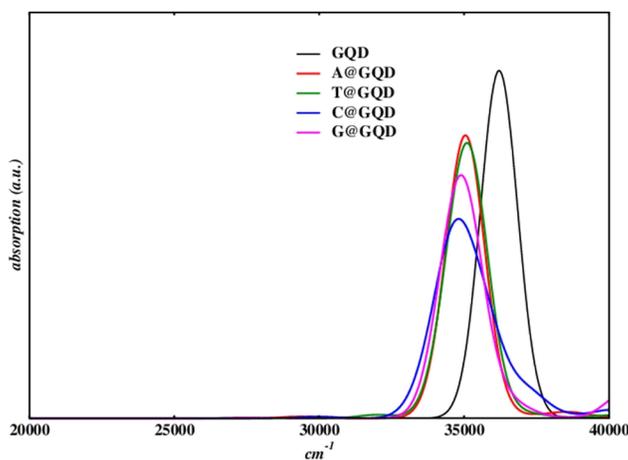

**FIGURE 2.** Calculated absorption spectra of pristine GQD and GQD with DNA nucleobase(s).

## IR Spectra

The IR-Vibrational spectroscopy is an interesting route that can be helpful in identification of DNA nucleobases by means of capturing the unique spectral features arising from DNA nucleobase. Experimentally, characteristic vibrational peaks can be captured using Fourier Transform Infrared Spectroscopy (FTIR). The calculated IR spectra for the pristine GQD and nucleobase@GQD systems are presented in figure 3.

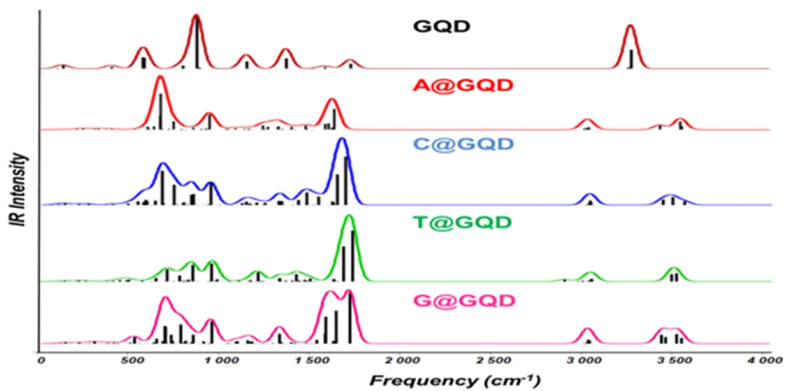

**FIGURE 3.** Calculated IR spectra of pristine GQD and GQD with DNA nucleobase(s).

The prominent peaks in calculated IR spectra have been identified on the basis of the major displacement vectors of the normal modes. There is a considerable modulation in the spectral bands of nucleobase@GQD systems as compared to pristine GQD. A marker peak below 1500 cm$^{-1}$ is found due to CH$_3$ normal mode for Thymine. Adenine exhibits a peak due to N-H normal modes in the vicinity of 1500 cm$^{-1}$. For the case of Cytosine peak near 1500 cm$^{-1}$ is sharper and broader for the case of Guanine. The sharp peak arises due to C=O normal modes in Cytosine while broader peak in the vicinity of 1500 cm$^{-1}$ is attributed to the closely spaced normal modes due to C=O and NH$_2$ functional groups in Guanine. Therefore, IR-vibrational spectroscopy provides various identifiable spectroscopic signatures in nucleobase@GQD systems near 1500 cm$^{-1}$ indicating the potential of GQD to probe the different nucleobases.

## CONCLUSIONS

Based on state-of-art electronic structure calculations, we find that DNA nucleobase are weakly physisorbed and exhibit the binding energy of the order of 0.1 eV. The GQD is more sensitive to Thymine and least sensitive to Adenine. An interaction of GQD with DNA nucleobase leads to modulation in electronic energy gap. The nucleobase origin of the new energy levels below the Fermi level is clearly captured with the help of Partial Density of States Analysis. The electronic band gap of pristine GQD reduces from 2.75 eV to 2.55 eV when guanine interacts with GQD. The UV/Vis spectra of conjugate systems allow a clear distinction of nucleobases from one another. The UV/Vis spectra shift towards slightly lower wavenumber as compared to pristine GQD.IR-vibration spectra for the composite systems exhibit distinct features. IR-vibrational spectra show unique spectral bands near wavenumber of 1500 cm$^{-1}$.

## ACKNOWLEDGEMENTS

Munish Sharma wishes to acknowledge the K2, high performance computational facility at Inter University Accelerator Centre (IUAC), New Delhi. We gratefully acknowledge SIESTA team and ORCA team for their code.

## REFERENCES

1. M. Sharma, A. Kumar, J. D. Sharma and P. Ahluwalia, Quantum Matter **5** (3), 315-318 (2016).
2. D. Raeyani, S. Shojaei, S. A. Kandjani and W. Wlodarski, Procedia Engineering **168**, 1312-1316 (2016).
3. G. Bharathi, D. Nataraj, S. Premkumar, M. Sowmiya, K. Senthilkumar, T. D. Thangadurai, O. Y. Khyzhun, M. Gupta, D. Phase and N. Patra, Scientific reports **7** (1), 10850 (2017).
4. Q.-L. Chen, C.-F. Wang and S. Chen, Journal of Materials Science **48** (6), 2352-2357 (2013).
5. S. S. Joo, J. Kim, S. S. Kang, S. Kim, S.-H. Choi and S. W. Hwang, Nanotechnology **25** (25), 255203 (2014).
6. X. Chen, Q. Jin, L. Wu, C. Tung and X. Tang, Angewandte Chemie International Edition **53** (46), 12542-12547 (2014).
7. D. Wang, J. F. Chen and L. Dai, Particle & Particle Systems Characterization **32** (5), 515-523 (2015).
8. M. K. Kumawat, M. Thakur, R. B. Gurung and R. Srivastava, ACS Sustainable Chemistry & Engineering **5** (2), 1382-1391 (2017).
9. C.-S. Chou, Y.-C. Wu and C.-H. Lin, Rsc Advances **4** (95), 52903-52910 (2014).
10. E. Artacho, E. Anglada, O. Diéguez, J. D. Gale, A. García, J. Junquera, R. M. Martin, P. Ordejón, J. M. Pruneda and D. Sánchez-Portal, Journal of Physics: Condensed Matter **20** (6), 064208 (2008).
11. T. Ahmed, S. Kilina, T. Das, J. T. Haraldsen, J. J. Rehr and A. V. Balatsky, Nano letters **12** (2), 927-931 (2012).
12. M. Sharma, Ashok Kumar, and P. K. Ahluwalia, RSC Advances **6** (65), 60223-60230. (2016).
13. H. Yoon, Y. H. Chang, S. H. Song, E. S. Lee, S. H. Jin, C. Park, J. Lee, B. H. Kim, H. J. Kang and Y. H. Kim, Advanced Materials **28** (26), 5255-5261 (2016).